\documentclass[aps,prd,twocolumn,nofootinbib,showpacs]{revtex4-2}

\usepackage{graphicx}
\usepackage[colorlinks=true, linkcolor=blue, citecolor=blue, urlcolor=blue]{hyperref}

\begin{document}

\title{Precise determination of the bottom-quark on-shell mass using its four-loop relation to the $\overline{\rm MS}$-scheme running mass}

\author{Shun-Yue Ma $^1$}

\author{Xu-Dong Huang $^{2}$}

\author{Xu-Chang Zheng $^1$}

\author{Xing-Gang Wu $^1$}

\affiliation{$^1$ Department of Physics, Chongqing Key Laboratory for Strongly Coupled Physics, Chongqing University, Chongqing 401331, P.R. China}
\affiliation{$^2$ College of Physics and Electronic Engineering, Chongqing Normal University, Chongqing 401331, P.R. China}

\date{\today}

\begin{abstract}

In this paper, we explore the properties of the bottom-quark on-shell mass ($M_b$) by using its relation to the $\overline{\rm MS}$ mass (${\overline m}_b$). At present, this $\overline{\rm MS}$-on-shell relation has been known up to four-loop QCD corrections, which however still has a $\sim 2\%$ scale uncertainty by taking the renormalization scale as ${\overline m}_b({\overline m}_b)$ and varying it within the usual range of $[{\overline m}_b({\overline m}_b)/2, 2 {\overline m}_b({\overline m}_b)]$. The principle of maximum conformality (PMC) has been adopted to achieve a more precise $\overline{\rm MS}$-on-shell relation by eliminating such scale uncertainty. As a step forward, we also estimate the magnitude of the uncalculated higher-order terms by using the Pad\'{e} approximation approach. Numerically, by using the $\overline{\rm MS}$ mass ${\overline m}_b({\overline m}_b)=4.183\pm0.007$ GeV as an input, our predicted value for the bottom-quark on-shell mass becomes $M_b\simeq 5.372^{+0.091}_{-0.075}$ GeV, where the uncertainty is the squared average of the ones caused by $\Delta \alpha_s(M_Z)$, $\Delta {\overline m}_b({\overline m}_b)$, and the estimated magnitude of the higher-order terms.

\hspace{-1em}\footnotesize{Keywords: bottom quark, on-shell mass, perturbative QCD, renormalization group}

\hspace{-1em}\footnotesize{PACS numbers: 12.38.Bx, 12.15.Ff, 12.10.Kt}
\end{abstract}

\maketitle

Quark masses are important parameters for the Quantum Chromodynamics (QCD) theory, which need to be renormalized in higher-order calculations. In perturbative QCD (pQCD) theory, two schemes are frequently adopted for renormalizing the quark masses, e.g. the on-shell (OS) scheme~\cite{Tarrach:1980up} and the modified minimal subtraction ($\overline{\rm MS}$) scheme~\cite{tHooft:1973mfk, Bardeen:1978yd}. The OS mass, also known as the pole mass, offers the advantage of being grounded in a physical definition which is gauge-parameter independent and scheme independent. It ensures that the inverse heavy-quark propagator exhibits a zero at the location of the pole mass to any order in the perturbative expansion. On the other hand, the $\overline{\rm MS}$ scheme focuses solely on removing the subtraction term $1/\epsilon+\ln(4\pi)-\gamma_{E}$ from the quantum corrections to the quark two-point function. And by combining this with the bare mass, one can derive the expression for the renormalized $\overline{\rm MS}$ mass. In high-energy processes, the $\overline{\rm MS}$ mass is preferred for its lack of intrinsic uncertainties. It has been found that for the high-energy processes involving the bottom quark, such as the $B$ meson decays, when their typical scales are lower than the bottom quark mass, the using of $\overline{\rm MS}$ mass becomes less suitable and the OS mass is usually adopted. Practically, the relation between the OS mass and the $\overline{\rm MS}$ mass that is affected by renormalons~\cite{Beneke:1994qe, Neubert:1994vb, Beneke:1998ui}, resulting in a perturbative series with poor convergence. Thus for precision tests of the Standard Model, accurate determination of the OS mass is important.

It is noted that the OS mass can be related to the $\overline{\rm MS}$ mass by using the perturbative relation between the bare quark mass ($m_{q,0}$) and the renormalized mass in either the OS or $\overline{\rm MS}$ scheme, where $q$ denotes the heavy charm, bottom, and top quark, respectively. For example, we have $m_{q,0}=Z^{\rm OS}_{m}M_q^{\rm OS}$ and $m_{q,0}=Z^{\overline{\rm MS}}_{m}{\overline m}_q(\mu_r)$, where $\mu_r$ is the renormalization scale. Here, $Z^{\rm OS}_{m}$ and $Z^{\overline{\rm MS}}_{m}$ represent the quark mass renormalization constants in the OS and $\overline{\rm MS}$ scheme, respectively. At present, the relation between the OS mass and the $\overline{\rm MS}$ mass, called as the $\overline{\rm MS}$-on-shell relation, has been calculated up to four-loop QCD corrections~\cite{Gray:1990yh, Chetyrkin:1999qi, Melnikov:2000qh, Jegerlehner:2002em, Jegerlehner:2003sp, Faisst:2004gn, Marquard:2007uj, Marquard:2015qpa, Marquard:2016dcn, AlamKhan:2023kgs}. Many attempts have been also finished to estimate the higher-loop contributions~\cite{Kataev:2018mob, Kataev:2018sjv, Kataev:2018gle}. Those improvements enable the possibility of precise determination of the bottom quark OS mass with the help of the experimentally fixed $\overline{\rm MS}$ mass. Both $\alpha_s$ and ${\overline m}_q$ are scale dependent, whose scale running behaviors are governed by the renormalization group equations (RGEs) that involve either the $\beta$-function~\cite{Politzer:1973fx, Politzer:1974fr, Gross:1973id, Gross:1973ju, Chetyrkin:2004mf, Baikov:2016tgj} or the quark mass anomalous dimension $\gamma_m$-function~\cite{Vermaseren:1997fq, Chetyrkin:1997dh, Baikov:2014qja}. Thus the crucial point for this determination is how to fix the precise values of $\alpha_s$ and the $\overline{\rm MS}$ running mass ${\overline m}_q$ simultaneously.

Practically, people usually uses the guessed renormalization scale and varies it within a certain range to estimate its uncertainty for a fixed-order pQCD series. This naive treatment leads to mismatches among the strong coupling constant and its expansion coefficients, which directly breaks the standard renormalization group invariance~\cite{Wu:2013ei, Wu:2014iba} and results in the conventional renormalization scale and scheme ambiguities. The effectiveness of this treatment depends heavily on the convergence of the pQCD series. Unfortunately, the bottom quark $\overline{\rm MS}$-on-shell relation exhibits poor convergence. The $\overline{\rm MS}$-on-shell relation up to four-loop QCD corrections still has a scale uncertainty about $100$ MeV~\cite{Marquard:2015qpa, Marquard:2016dcn}, which significantly exceeds the current uncertainty of the $\overline{\rm MS}$ mass of the bottom quark issued by the Particle Data Group~\cite{ParticleDataGroup:2024cfk}, ${\overline m}_b({\overline m}_b)=4.183\pm0.007$ GeV.

In the process of renormalization, a renormalization scale needs to be introduced. For the original truncated perturbative series, if the renormalization scale is chosen improperly, the perturbative series will depend on the choice of renormalization scale and renormalization scheme. However, with methods such as the principle of maximum conformality (PMC)~\cite{Brodsky:2011ta, Brodsky:2011ig, Brodsky:2012rj, Mojaza:2012mf, Brodsky:2013vpa, Shen:2017pdu}, the improved perturbative series leads to a prediction without renormalization scale ambiguity in the fixed-order. The PMC offers a systematic approach for determining the correct value of $\alpha_s$ by using the RGE-involved $\{\beta_i\}$-terms of the pQCD series. After using those $\{\beta_i\}$-terms, the initial pQCD series changes into a newly scheme-independent conformal series. It has been found that the resultant PMC series is independent of any choice of renormalization scale, and the scale-invariant PMC series is also valuable for estimating the contributions of uncalculated higher-order (UHO) terms~\cite{Du:2018dma}. The comprehensive exploration of the PMC can be found in the review articles~\cite{Wu:2019mky, DiGiustino:2023jiq}.

Recently, an improved PMC scale-setting procedure has been proposed in Ref.\cite{Huang:2022rij}. This approach simultaneously determines the correct magnitudes of $\alpha_s$ and the quark mass ${\overline m}_q$ by utilizing the RGEs for determining the magnitudes of the running coupling $\alpha_s$ and the running mass under the same scheme such as the $\overline{\rm MS}$ scheme. Upon implementing the new scale-setting procedure to the $\overline{\rm MS}$-on-shell relation, the renormalization scale ambiguity inherent in the pQCD series is effectively eliminated. Additionally, the renormalon terms associated with the $\beta$-function and $\gamma_m$-function of the pQCD series can also be removed. In Ref.\cite{Huang:2022rij}, we have made a detained discussion on the top-quark pole mass. In this Letter, we intend to utilize the PMC scale-setting approach to determine the bottom-quark OS mass. Furthermore, the Pad\'{e} approximation approach (PAA)~\cite{Basdevant:1972fe, Samuel:1992qg, Samuel:1995jc}, which offers a systematic method for approximating a finite perturbative series into an analytic function, will be employed to estimate the $(n+1)_{\rm th}$-order coefficient by incorporating all known coefficients up to order $n$. The PAA has been shown to be effective both in predicting unknown higher-order coefficients and in summing the perturbative series.

The relationship between the bottom-quark $\overline{\rm MS}$ quark mass and its OS quark mass can be expressed as
\begin{eqnarray}
&&\frac{M_b}{{\overline m}_b(\mu_r)}=\frac{Z_m^{\overline{\rm MS}}}{Z_m^{\rm OS}}=\sum_{n\geq0}a^n_s(\mu_r) c^{(n)}_m(\mu_r),
\end{eqnarray}
where $a_s=\alpha_s/(4\pi)$, $c^{(0)}_m(\mu_r)=1$ and $c^{(n)}_m(\mu_r)$ is a function of $\ln(\mu_r^2/{\overline m}_b^2(\mu_r))$. Subsequently, the determination of the bottom-quark OS mass can be achieved using the following relationship:
\begin{eqnarray}
M_b|_{\rm conv.} &=&{\overline m}_b(\mu_r)\Big\{1+{\mathcal C}_1(\mu_r)a_s(\mu_r) + {\mathcal C}_2(\mu_r)a_s^2(\mu_r)\nonumber \\
&&+{\mathcal C}_3(\mu_r)a_s^3(\mu_r)+ {\mathcal C}_4(\mu_r)a_s^4(\mu_r)+\mathcal{O}(a^5_s)\Big\}, \label{relation}
\end{eqnarray}
where conv. stands for the initial pQCD series given under the $\overline{\rm MS}$ scheme. As mentioned above, the expansion coefficients ${\mathcal C}_{i}$ have been known up to N$^4$LO-level, which need to be transformed as the $\beta$-series so as to fix the correct values of $\alpha_s$ and ${\overline m}_b$. This transformation can be done by using the general QCD degeneracy relations, and the results are given in Ref.\cite{Huang:2022rij}. Then, by applying the PMC scale-setting procedures, Eq.(\ref{relation}) can be transformed into the following conformal series:
\begin{eqnarray}
M_b|_{\rm PMC}&=&{\overline m}_b(Q_*)\Big\{1+ r_{1,0}a_s(Q_*) + r_{2,0}a_s^2(Q_*)\nonumber \\
&&+r_{3,0}a_s^3(Q_*)+ r_{4,0}a_s^4(Q_*)+\mathcal{O}(a^5_s)\Big\},  \label{PMCrelation}
\end{eqnarray}
where $r_{i,0}$ are conformal coefficients. Here $Q_*$ represents the PMC scale, whose logarithmic form $\ln(Q_{*}^{2}/{\overline m}_b^2(Q_*))$ can be represented as a power series in $a_s(Q_*)$:
\begin{eqnarray}
	\ln\frac{Q^2_*}{{\overline m}_b^2(Q_*)}=\sum_{i=0}^{n} S_{i} a^i_s(Q_*),  \label{qstar1}
\end{eqnarray}
where the coefficients $S_i$ can be determined up to next-to-next-to-leading log (NNLL) accuracy~\cite{Huang:2022rij} by using the given four-loop pQCD series. It is found that Eq.(\ref{qstar1}) is independent to the choice of renormalization scale. This property ensures that both the running mass ${\overline m}_b$ and the running coupling constant $\alpha_s$ are concurrently determined. By matching the $\mu_r$-independent conformal coefficients $r_{i,0}$, the resulting PMC series becomes devoid of the conventional renormalization scale ambiguity.

We are now ready to calculate the bottom-quark OS mass $M_b$ through its perturbative relation to the $\overline{\rm MS}$ mass. To do the numerical computations, we adopt $\alpha_s(M_Z)=0.1180\pm0.0009$ and ${\overline m}_b({\overline m}_b)=4.183\pm0.007$ GeV~\cite{ParticleDataGroup:2024cfk}. The scale running of $\alpha_s(\mu_r)$ is calculated by using the package RunDec~\cite{Herren:2017osy}.

\begin{figure}[htb]
\includegraphics[width=0.48\textwidth]{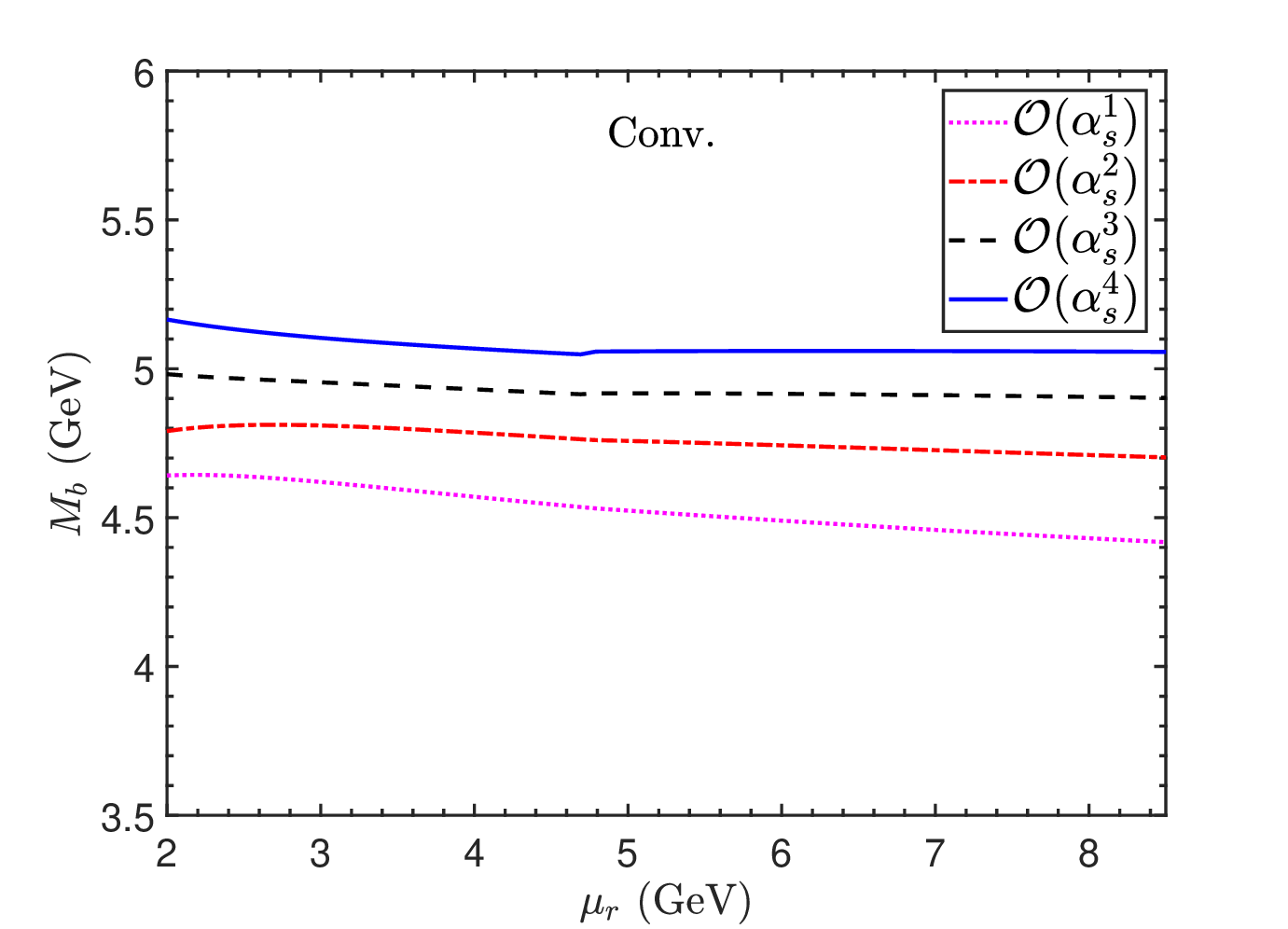}
\caption{The bottom-quark OS mass $M_b$ versus the renormalization scale ($\mu_r$) under conventional scale-setting approach up to various perturbative orders.} \label{Convurdepen}
\end{figure}

Using Eq.(\ref{relation}) and setting all the input parameters to be their central values, we present the bottom-quark OS mass $M_b$ under conventional scale-setting approach in FIG.~\ref{Convurdepen}. FIG.~\ref{Convurdepen} shows how the bottom-quark OS mass $M_b$ changes with the renormalization scale and demonstrates that the conventional renormalization scale dependence diminishes as more loop terms have been incorporated. Numerically, we have
\begin{eqnarray}
M_b|_{\rm Conv.}&=&4.183^{+0.782}_{-0.111} +0.399^{-0.584}_{+0.062} + 0.199^{-0.162}_{+0.027} \nonumber \\
&&+ 0.145^{+0.023}_{+0.007} + 0.136^{+0.036}_{-0.001}\nonumber \\
&=&5.062^{+0.095}_{-0.016}~(\rm GeV), \label{perturbativeseries}
\end{eqnarray}
whose central values are for $\mu_r={\overline m}_b({\overline m}_b)$, and the uncertainties are for $\mu_r \in [{\overline m}_b({\overline m}_b)/2, 2{\overline m}_b({\overline m}_b)]$.The relative magnitudes of the leading-order terms (LO): the next-to-leading-order terms (NLO): the next-to-next-to-leading-order terms (N$^2$LO): the next-to-next-to-next-to-leading-order terms (N$^3$LO): the next-to-next-to-next-to-next-to-leading-order terms (N$^4$LO) are approximately 1: $9.5\%$: $4.8\%$: $3.5\%$: $3.3\%$ for the case of $\mu_r={\overline m}_b({\overline m}_b)$.  Eq.(\ref{perturbativeseries}) shows the magnitudes of each loop terms are highly scale dependent, and the perturbative behavior of the whole series is different for different scale choices. Within this scale range, the absolute scale uncertainties are about $21\%$, $162\%$, $95\%$, $16\%$, and $27\%$ for the LO, the NLO, the N$^2$LO, the N$^3$LO, and the N$^4$LO terms, respectively. The overall scale uncertainty of the four-loop prediction of $M_b$ becomes $\sim 2.2\%$ due to the large cancellation of scale dependence among different orders.

\begin{figure}[htb]
\includegraphics[width=0.48\textwidth]{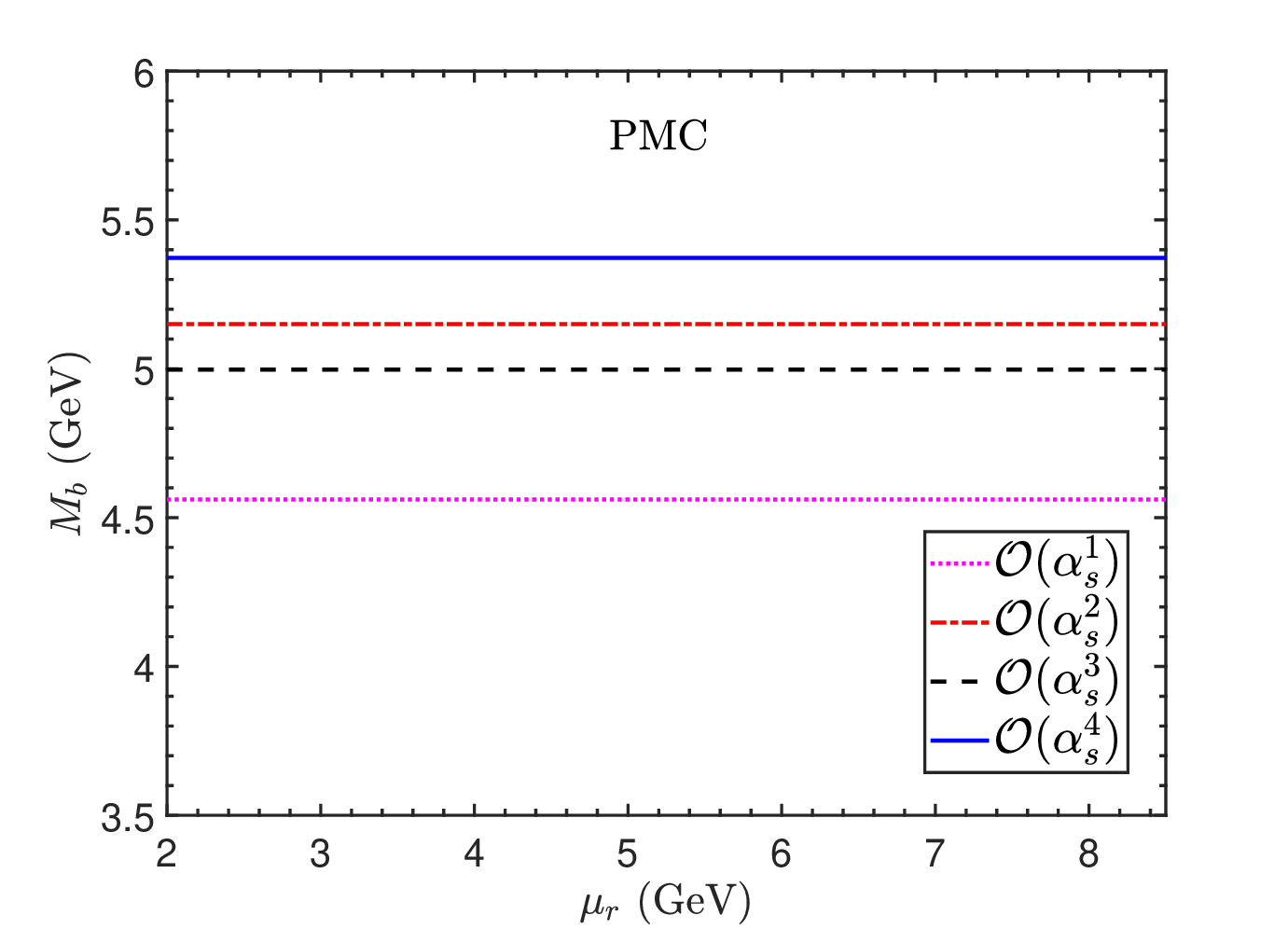}
\caption{The bottom-quark OS mass $M_b$ versus the renormalization scale ($\mu_r$) under PMC single-scale approach up to different perturbative orders.} \label{pmcurdepen}
\end{figure}

Similarly, using Eq.(\ref{PMCrelation}), we present $M_b$ under the PMC scale-setting approach in FIG.~\ref{pmcurdepen}. The PMC scale $Q_*$ can be fixed up to N$^2$LL accuracy by using Eq.(\ref{qstar1}), i.e.,
\begin{eqnarray}
\ln\frac{Q^2_*}{{\overline m}_b^2(Q_*)}&=&-72.8957 a_s(Q_*)+137.61 a^2_s(Q_*)\nonumber \\
&&-17177.2 a^3_s(Q_*),
\end{eqnarray}
which is independent to any choice of $\mu_r$ and leads to $Q_*=1.917$ GeV. The flat lines in FIG.~\ref{pmcurdepen} indicates that the PMC prediction is devoid of renormalization scale ambiguity at any fixed order. Numerically, we have
\begin{eqnarray}
M_b|_{\rm PMC}&=&5.100 + 0.667 - 0.439 - 0.153 + 0.197 \nonumber \\
&=&5.372~(\rm GeV). \label{PMCseries}
\end{eqnarray}
It shows that the relative importance of the LO: the NLO: the N$^2$LO: the N$^3$LO: the N$^4$LO terms in the PMC series is 1: $13.0\%$: $-8.6\%$: $-3.0\%$: $3.9\%$.

It has been found that for the case of top quark~\cite{Huang:2022rij}, whose $\alpha_s(\overline{m}_t)\sim 0.11$ and $\alpha_s(Q^*_t=123.3~{\rm GeV})\sim 0.11$, there is good perturbative behavior for both the conventional and PMC series. However, for the present case of bottom quark, whose $\alpha_s(\overline{m}_b)\sim 0.22$ and $\alpha_s(Q^*_b=1.917~{\rm GeV})\sim 0.31$, the $\alpha_s$ power suppression fails to counterbalance the influence of the substantial numerical coefficients even after applying the PMC, e.g. Eq.(\ref{perturbativeseries}) shows that the relative importance of the N$^2$LO: N$^3$LO: N$^4$LO terms for conventional series is 1: $73\%$: $68\%$ for $\mu_r={\overline m}_b({\overline m}_b)$, 1: $454\%$: $465\%$ for $\mu_r={\overline m}_b({\overline m}_b)/2$, and 1: $67\%$: $60\%$ for $\mu_r=2 {\overline m}_b({\overline m}_b)$, respectively; and Eq.(\ref{PMCseries}) shows that such relative importance changes to 1: $35\%$: $45\%$ for the scale-invariant PMC series. At present, the relatively large magnitude of the N$^4$LO-terms indicates that the magnitude of the N$^4$LO conformal coefficients, which are unrelated to the RGE-involved $\beta$-terms or the quark mass anomalous dimension involved $\gamma_m$-terms, is large. However such scale-invariant perturbative behavior can be treated as the intrinsic perturbative behavior of the $\overline{\rm MS}$-on-shell relation. By properly choosing the scale, the perturbative behavior of conventional series will be close to the PMC one. Thus for those cases, a proper scale-setting approach to achieve a scale-invariant series is very important. Moreover, to compare with the N$^2$LO-terms and N$^3$LO-terms, the sizable magnitude of the N$^4$LO-terms indicates the importance of knowing whether the UHO-terms can give sizable contributions and present the wanted convergent behavior. For the purpose, we adopt the PAA to estimate the magnitude of the UHO contributions.

\begin{table}[htb]
\begin{tabular}{cccc}
\hline
      \raisebox {0ex}[0pt]{~~~~}
         & ~~$m=3$~~    & ~~$m=4$~~    & ~~$m=5$~~  \\
\hline
      \raisebox {0ex}[0pt]{~~$M_b|_{\rm PAA}^{N/M}$~~}
       & ~~$[0/1] 4.976$~~ & ~~$[0/2] 5.216$~~ & ~~$[0/3] 5.586$~~ \\
       & ~~$-$~~ & ~~$[1/1] 5.317$~~ & ~~$[1/2] 3.441$~~ \\
       & ~~$-$~~ & ~~$-$~~ & ~~$[2/1] 7.090$~~ \\
\hline
\end{tabular}
\caption{The $m_{\rm th}$-loop PAA predictions $M_b|_{\rm PAA}^{N/M}$ (in unit: \rm GeV) for the conventional series with $\mu_r={\overline m}_b({\overline m}_b)$, where $N+M=m-2$. }  \label{convpaa}
\end{table}

\begin{table}[htb]
\centering
\begin{tabular}{cccc}
\hline
      \raisebox {0ex}[0pt]{~~~~}
              & ~~$m=3$~~    & ~~$m=4$~~    & ~~$m=5$~~  \\
\hline
      \raisebox {0ex}[0pt]{~~$M_b|_{\rm PAA}^{N/M}$~~}
       & ~~$[0/1]$ 5.295~~ & ~~$[0/2] 5.161$~~ & ~~$[0/3] 5.355$~~ \\
       & ~~$-$~~ & ~~$[1/1] 4.939$~~ & ~~$[1/2] 5.335$~~ \\
       & ~~$-$~~ & ~~$-$~~ & ~~$[2/1] 5.261$~~ \\
\hline
\end{tabular}
\caption{The $m_{\rm th}$-loop PAA predictions $M_b|_{\rm PAA}^{N/M}$ (in unit: \rm GeV) for the PMC series, where $N+M=m-2$.  }  \label{pmcpaa}
\end{table}

The PAA is a kind of resummation to create an appropriate generating function such as the fractional generating function; and it offers a systematic way for approximating a finite perturbative series into an analytic function. For a given pQCD series that can be written as $\rho(Q)=\sum^n\limits_{i=1} C_i a_s^i$, its $[N/M]$-type fractional generating function is defined as~\cite{Basdevant:1972fe, Samuel:1992qg, Samuel:1995jc},
\begin{eqnarray}
\rho^{N/M}(Q)&=&a_s\times\frac{b_0+b_1 a_s+\cdots+b_N a_s^N}{1+c_1 a_s+\cdots+c_M a_s^M},
\end{eqnarray}
where $N$ and $M$ are integers, $N\geq0$, $M\geq1$, $N+M+1=n$. The coefficients $b_{i\in[0,N]}$ and $c_{i\in[1,M]}$ can be fixed by requiring the coefficients $C_{i\in[1,n]}$ defined in the following expansion series 
\begin{eqnarray}
\rho^{N/M}(Q)&=&\sum^n_{i=1} C_i a_s^i
\end{eqnarray}
to be the same. The coefficients $C_i={\mathcal C}_i$ and $C_i=r_{i,0}$ for $i \leq n$ for the conventional series and PMC conformal series, respectively. Using Eqs.(\ref{relation}, \ref{PMCrelation}), the $\overline{\rm MS}$-on-shell relation up to $m_{\rm th}$-loop level can be written as the following form,
\begin{displaymath}
M_b|_{\rm PAA}^{N/M}={\overline m}_b(\mu)[1+\rho^{N/M}(\mu)],
\end{displaymath}
where $N+M=m-2$ due to $n=m-1$\footnote{The perturbative series of $M_b|_{\rm PAA}^{N/M}$ is one-order higher than the perturbative series of $\rho^{N/M}(\mu)$, since the perturbative series of $M_b|_{\rm PAA}^{N/M}$ includes $\alpha_s^0$-order terms after factoring ${\overline m}_b(\mu)$. $m$ and $n$ denote the accuracy of the $M_b|_{\rm PAA}^{N/M}$ and $\rho^{N/M}(\mu)$, respectively.}, $\mu=\mu_r$ for conventional series and $\mu=Q_*$ for the PMC series. The PAA works for $m\geq 3$. Following the standard PAA procedures described in detail in Ref.\cite{Du:2018dma}, we obtain the different types of PAA predictions by using the known two-, three-, and four-loop pQCD series and put them in Table~\ref{convpaa} and Table~\ref{pmcpaa}.

Due to the presence of divergent renormalon terms associated with the $\beta$-function and $\gamma_m$-function in each loop term, the PAA prediction derived from the conventional series exhibits significant uncertainty. The PAA prediction is generally $[N/M]$-type dependent, which provides the systematic error of the PAA approach. More explicitly, Table~\ref{convpaa} shows the relative magnitudes of the allowable types of PAA predictions are $M_b|_{\rm PAA}^{m=3}: M_b|_{\rm PAA}^{m=4}: M_b|_{\rm PAA}^{m=5}= 1: 1.05\sim1.07: 0.69\sim1.42$ for the conventional series, respectively. On the contrary, the PMC conformal series, which is free of renormalon terms associated with the $\beta$-function and $\gamma_m$-function, can be a more reliable foundation for predicting UHO contribution. Table~\ref{pmcpaa} shows the relative magnitudes of the allowable types of PAA predictions are $M_b|_{\rm PAA}^{m=3}: M_b|_{\rm PAA}^{m=4}: M_b|_{\rm PAA}^{m=5}= 1: 0.93\sim0.97: 0.99\sim1.01$ for the PMC series, respectively. It has been found that the $[0/n-1]$ or $[0/m-2]$-type PAA predictions are self-consistent for the PMC method itself~\cite{Du:2018dma}, which agrees with the GM-L scale-setting procedure~\cite{GellMann:1954fq} to obtain scale-independent perturbative QED predictions. Moreover, one may also observe that the $[0/n-1]$-type PAA predictions under different orders exhibits better stability than other types. So, we adopt the $[0/n-1]$-type PAA predictions as an estimate of the UHO contributions, i.e.,
\begin{eqnarray}
\Delta M_b|_{\rm Conv.}^{\rm High~order}&=&\pm\Big|M_b|_{\rm PAA, Conv.}^{[0/3]}-M_b|_{\rm Conv.}\Big| \nonumber \\
&=&\pm0.524~(\rm GeV), \\
\Delta M_b|_{\rm PMC}^{\rm High~order}&=&\pm\Big|M_b|_{\rm PAA, PMC}^{[0/3]}-M_b|_{\rm PMC}\Big| \nonumber \\
&=&\pm0.017~(\rm GeV).
\end{eqnarray}
Due to the elimination of divergent renormalon terms, the PMC series predicts a much smaller uncertainty from the UHO-terms. And it is also found that the PAA works better when more terms have been known. We define a ratio, $\kappa_i=\Big|(M_b|_{\rm PAA}^{[0/n-1]}-M_b^{\mathcal{O}(\alpha_s^n)})/M_b^{\mathcal{O}(\alpha_s^n)}\Big|$ with $n=(2,3,4)$, respectively, which shows more explicitly how more loop terms affect the accuracy of the PAA prediction. For conventional series, we have $\kappa_2:\kappa_3:\kappa_4 =0.041: 0.059: 0.104$; and for the PMC conformal series, we have $\kappa_2:\kappa_3:\kappa_4 =0.028: 0.033: 0.003$.

In addition to the uncertainties due to UHO-terms, there are also uncertainties from the $\Delta\alpha_s(M_Z)$ and $\Delta{\overline m}_b({\overline m}_b)$. Using $\alpha_s(M_Z)=0.1180\pm0.0009$~\cite{ParticleDataGroup:2024cfk} as an estimate, we obtain
\begin{eqnarray}
\Delta M_b|_{\rm Conv.}^{\Delta\alpha_s(M_Z)}&=&(^{+0.028}_{-0.026})~(\rm GeV), \\
\Delta M_b|_{\rm PMC}^{\Delta\alpha_s(M_Z)}&=&(^{+0.089}_{-0.073})~(\rm GeV).
\end{eqnarray}
Using the RGE to fix the correct magnitude of $\alpha_s$, the PMC series thus depends heavily on the precise $\alpha_s$ running behavior. The more sensitivity of the PMC series on the value of $\alpha_s(M_Z)$ makes it inversely be a better platform to fix the reference point value from comparison of experimental data~\cite{Shen:2023qgz}.

Regarding the uncertainty arising from the choice of the bottom-quark $\overline{\rm MS}$ mass, $\Delta{\overline m}_b({\overline m}_b)=\pm0.007$ GeV, we obtain
\begin{eqnarray}
\Delta M_b|_{\rm Conv.}^{\Delta{\overline m}_b({\overline m}_b)}&=&(^{+0.008}_{-0.007})~(\rm GeV), \\
\Delta M_b|_{\rm PMC}^{\Delta{\overline m}_b({\overline m}_b)}&=&(^{+0.007}_{-0.005})~(\rm GeV).
\end{eqnarray}
This indicates that the bottom-quark OS mass could depend almost linearly on its $\overline{\rm MS}$ mass, since the uncertainty is at the same order of ${\cal O}(\Delta{\overline m}_b({\overline m}_b))$.

In summary, we have determined the bottom-quark OS mass using the four-loop $\overline{\rm MS}$-on-shell relation in conjunction with the newly suggested PMC approach, which determines the correct magnitudes of the $\alpha_s$ and the $\overline{\rm MS}$-running mass simultaneously by using the $\beta$-function and $\gamma_m$-function of the pQCD series. Taking the bottom-quark $\overline{\rm MS}$ mass ${\overline m}_b({\overline m}_b)=4.183\pm0.007$ GeV as an input, we have derived a precise bottom-quark OS mass:
\begin{eqnarray}
M_b|_{\rm Conv.}&=&5.062^{+0.533}_{-0.525}~(\rm GeV), \\
M_b|_{\rm PMC}&=&5.372^{+0.091}_{-0.075}~(\rm GeV),
\end{eqnarray}
where the uncertainties stem from the mean square of those originating from $\Delta M_b|^{\text{High~order}}$, $\Delta M_b|^{\Delta\alpha_s(M_Z)}$, and $\Delta M_b|_{\rm Conv.}^{\Delta{\overline m}_b({\overline m}_b)}$, respectively. It is important to note that the conventional prediction still exhibits renormalization scale uncertainty, which arises from varying $\mu_r$ within the range $\mu_r \in[{\overline m}_b({\overline m}_b)/2, 2{\overline m}_b({\overline m}_b)]$.

The accuracy of the pQCD predictions within the framework of the $\overline{\rm MS}$ running mass scheme is critically dependent on the precise determination of $\alpha_s$ and ${\overline m}_q$. With the implementation of the PMC approach, the accurate values of the effective $\alpha_s$ and ${\overline m}_q$ can be ascertained. This results in a more convergent pQCD series, thereby reducing uncertainties and promoting the attainment of a reliable and precise pQCD prediction.

{\it Acknowledgments.} This work was supported in part by the National Natural Science Foundation of China under Grant No.12175025, No.12247129, and No.12347101, by the Graduate Research and Innovation Foundation of Chongqing, China under Grant No.ydstd1912, and by the Foundation of Chongqing Normal University under Grant No.24XLB015.

\end{document}